\documentclass[9pt,arxiv]{lapreprint}
\usepackage[version=4]{mhchem} 
\usepackage{siunitx}    
\usepackage{pdflscape}  
\usepackage{rotating}   
\usepackage{textgreek}  
\usepackage{gensymb}    
\usepackage[misc]{ifsym} 
\usepackage{listings}   
\usepackage{colortbl}   
\usepackage{tabularx}   
\usepackage{longtable}  
\usepackage{subcaption}
\usepackage{multirow}
\usepackage{snotez}     
\usepackage{csquotes}   

\DeclareSIUnit\Molar{M}

\usepackage[			
	backend=bibtex,      
    style=authoryear,   
    natbib=true,
	hyperref=true,	    
	alldates=year      
]{biblatex}

\usepackage{libertine}

\AtEveryBibitem{
    \clearfield{urlyear}
    \clearfield{urlmonth}
    \clearlist{language}
}

\usepackage{amsthm,amsmath}
\usepackage{booktabs}
\usepackage{graphicx}
\usepackage{tabularx}
\usepackage{url}
\usepackage{subcaption}

\title{Attitudes Towards Migration in a COVID-19 Context: Testing a Behavioral Immune System Hypothesis with Twitter Data}

\leadauthor{Yerka Freire-Vidal}
\shorttitle{Testing a Behavioral Immune System Hypothesis with Twitter Data}

\author[1 \Letter]{Yerka Freire-Vidal}
\author[2]{Gabriela Fajardo}
\author[2]{Carlos Rodr\'iguez-Sickert}
\author[3,4]{Eduardo Graells-Garrido}
\author[1]{Jos\'e Antonio Mu\~noz-Reyes}
\author[1]{Oriana Figueroa}

\affil[1]{Social Complexity Research Center, Universidad del Desarrollo, Chile}
\affil[2]{Facultad de Administraci\'on y Econom\'ia, Universidad de Santiago de Chile, Chile}
\affil[3]{Department of Computer Science, University of Chile, Chile}
\affil[4]{National Center for Artificial Intelligence (CENIA), Chile}

\metadata[]{\Letter\hspace{.5ex} For correspondence}{y.freire@udd.cl (YFV)}
\metadata[]{Funding}{E. Graells-Garrido was partially funded by National Center for Artificial Intelligence CENIA, Basal ANID FB210017.}
\metadata[]{Competing interests}{The author declare no competing interests.}
\begin{document}

\maketitle

\begin{abstract}
The COVID-19 outbreak implied many changes in the daily life of most of the world's population for a long time, prompting severe restrictions on sociality. The Behavioral Immune System (BIS) suggests that when facing pathogens, a psychological mechanism would be activated that, among other things, would generate an increase in prejudice and discrimination towards marginalized groups, including immigrants.
This study aimed to test if people tend to enhance their rejection of minorities and foreign groups under the threat of contagious diseases, using the users' attitudes towards migrants in Twitter data from Chile, for pre-pandemic and pandemic contexts. Our results only partially support the BIS hypothesis, since threatened users increased their tweet production in the pandemic period, compared to empathetic users, but the latter grew in number and also increased the reach of their tweets between the two periods. We also found differences in the use of language between these types of users. Alternative explanations for these results may be context-dependent.
\end{abstract}


\section{Introduction}

The COVID-19 pandemic has precipitated a global change in social dynamics, intensifying the activation of the Behavioral Immune System (BIS), a psychological defense mechanism developed to detect and avoid infectious diseases through the constriction of different aspects of sociality \parencite{schaller2006parasites, schaller2011behavioral, mortensen2010infection}. The BIS is interrelated with the emotion of disgust, causes cognitive and behavioral modifications to mitigate pathogen transmission~\parencite{tybur2009microbes, tybur2013disgust}. These modifications extend to altered perceptions and behaviors towards others, often exacerbating prejudice and discrimination against marginalized groups, particularly migrants~\parencite{mortensen2010infection}.

In particular, the BIS and the Pathogen Disgusting System (PDS), as discussed in evolutionary psychology, share functional similarities~\parencite{tybur2009microbes, tybur2013disgust, lieberman2014behavioral}. Both systems are responses to the evolutionary pressures of a disease, generating proactive avoidance behaviors toward possible sources of infection. This convergence suggests a fundamental mechanism underlying social responses to a threat, which extends to xenophobic attitudes~\parencite{sarolidou2020people}.

The existing literature predominantly relies on survey methodologies to correlate xenophobia with pathogen avoidance, leaving a gap in empirical evidence from behavioral data~\parencite{navarrete2006disease}. Our research aims to address this by analyzing Twitter discourse during the COVID-19 pandemic, testing the hypothesis that elevated perceptions of vulnerability to disease are associated with greater negative sentiment toward foreigners. As such, this work contributes to the understanding of the BIS and the PDS, and also to the broader implications of disease avoidance mechanisms on social attitudes in a pandemic context.
\textcite{hruschka2013institutions}, in their experimental design, found no evidence to support the BIS hypothesis. Instead, they found that government efficacy is more relevant in explaining in-group favoritism (related to xenophobic tendencies) than disease avoidance. Such discriminatory and prejudiced behavior in times of pandemics is nothing new. Throughout history, it has emerged as an increase of threat and fear in populations. For example, there was persecution of Jews during the Black Death in the 14th century~\parencite{link2006stigma}. Later in the 1980s, lesbian, gay, bisexual, and transgender communities suffered great stigmatization and social exclusion during the HIV outbreak~\parencite{berger2001measuring}, and more recently, the Ebola outbreak was labeled as the ``African disease''~\parencite{davtyan2014addressing}. During the COVID-19 pandemic, increased moral disgust has been demonstrated during periods of high pathogen threat, with individual responses influenced by age and trait anxiety, reflecting deep-seated psychological predispositions rather than just the immediate context~\parencite{schwambergova2023pandemic}. Nevertheless, a pandemic outbreak is not necessary to explain in-group favoritism and the more intense xenophobic manifestations in several populations~\parencite{ogunyemi2020psychological, masuda2015evolutionary}. Therefore, we find it interesting to discuss and contrast the theoretical models designed to justify discriminatory behavior in a pandemic context.

Without considering the presence of a pandemic outbreak, many studies have focused on finding the factors that influence attitudes toward migration, because of the consequences that they have on social cohesion~\parencite{hopkins2010politicized, jolly2014xenophobia, kopstein2009does, sniderman2000outsider}, and on migrants’ marginalization. This affects not only individual psychological and socioeconomic factors~\parencite{burns2000economic, scheve2001labor} but the relationship between migrant population and locals, with emerging conflicts that in many cases lead to racism and xenophobia~\parencite{brown2003teammates, herek1986instrumentality, pettigrew2006meta}. One relevant theoretical framework developed to understand such attitudes is the ``Integrated Threat Theory''~\parencite{nelson2009handbook, stephan2013integrated}. This theory proposes that, under certain conditions, contact between local and immigrant populations generates perceptions of threat and risk, fostering discriminatory and exclusionary attitudes on the local population. The concept of ``contact'' is not limited to physical contact, but also encompasses indirect, imaginary, or virtual contact~\parencite{amichai2006contact, crisp2009can, dovidio2011improving, white2012dual}; and the type of perceived threat has been related mainly to competition for jobs, public services, and economic factors in general~\parencite{burns2000economic, scheve2001labor}.

Currently, the number of international migrants worldwide has reached 281 million, equivalent to 3.6\% of the world population~\parencite{OIMmigrantpop}. In Chile, this number amounts to 1.7 million people (7.2\% of its population)~\parencite{Cepal}. Our research seeks to measure attitudes towards migrants before and during the most critical moments of the pandemic in Chile, using Twitter~\parencite{freire2021framework}. In this way, we could shed light on how the BIS is operating in a society that has seen its immigrant population grow dramatically in recent years and has been jointly affected by the global COVID-19 pandemic. We have baseline measures before the pandemic outbreak, which gives us the possibility to isolate the effect of the behavioral immune system over attitudes towards immigrants as a more specific case than what is explained by the ``threat theory''. 

Our proposal uses the information that people post on Twitter about migration as a reflection of their attitudes toward this issue since they express their ideas and opinions voluntarily. Studies that have used Twitter reveal socio-cultural characteristics of users or societies, including the influence of culture in personal actions, political polarization~\parencite{garcia2013cultural}, personality traits~\parencite{quercia2011our}, personality differences between democrats and republicans~\parencite{sylwester2015twitter}, but also the level of integration of immigrants in a city~\parencite{lamanna2018immigrant}, and attitudes in response to triggering events such as terrorist attacks~\parencite{darwish2018predicting}. In this way, our proposal seeks to complement traditional methods for this topic, such as qualitative case studies or context-specific surveys for each migration context~\parencite{hainmueller2014public, o2006determinants, poutvaara2018bitterness}. Thus, we contribute not only with a dynamic approach to attitudes but also through an easily accessible and low-cost data source.

Particularly, the aim of this study is to test one of the core statements of the Behavioral Immune System (BIS): that people tend to enhance their rejection of minorities and foreign groups under the threat of contagious diseases. Specifically, we expect an increase in the number of Twitter users with threatened attitudes (or a decrease in the number of users with positive or empathetic attitudes), between the periods before and after the COVID-19 outbreak (called the pre- and post-pandemic periods, hereafter). Similarly, the number of threatened tweets should also increase (or the number of positive or empathetic tweets should decrease). We also expect an increase in the number of retweets from the group of users with threatened attitudes between pre-pandemic and pandemic contexts (or a decrease in the number of retweets from the positive or empathetic group of users), which is understood as a greater diffusion and scope of its negativity. Lastly, we expect the language use of the threatened user group to reflect a greater concern for the threat of COVID-19 contagion, i.e., they should use COVID-19-related words more frequently than positive or empathetic groups, especially in the pandemic context.

\section{Methodology}
\label{sec:metodo}

\subsection{Dataset Description}

Our dataset was extracted from the Twitter platform. Twitter is a micro-blogging platform in which each user with an account can create a user name, inform a location, and include a brief personal description, an image, and a URL in their profiles; this information can be fictitious or real. In addition, users can post tweets (texts of a maximum of 280 characters at the time of data collection), retweets (share other users' Tweets on their account), mention others in their tweets using an identifier (e.g., @username), and quote others' tweets or add comments to them. Twitter users can follow other users and view their tweets on their own timelines. 

We used a system designed to collect Chilean tweets~\parencite{graells2016encouraging} which connects to the Twitter Streaming API. Then, to download those referring to migration we used a list of keywords related to the topic. The query parameters included words such as ``migration (migraci\'on)'', ``immigration (inmigraci\'on)'', ``migrant  (migrante)'', ``immigrant (inmigrante'', ``foreigner (extranjero)'', ``borders (fronteras)'', etc., and countries of origin with their respective demonyms (see the complete list in the Appendix~\ref{appendix:keywords}; note that the original list of keywords is in Spanish, here we report in English for clarity). Lastly, we pre-processed the data to eliminate noise topics such as bird migration, migration in other countries, etc.

\begin{figure*}[t]
  \subfloat{
   \label{f:tweets_pre}
    \includegraphics[width=0.5\textwidth]{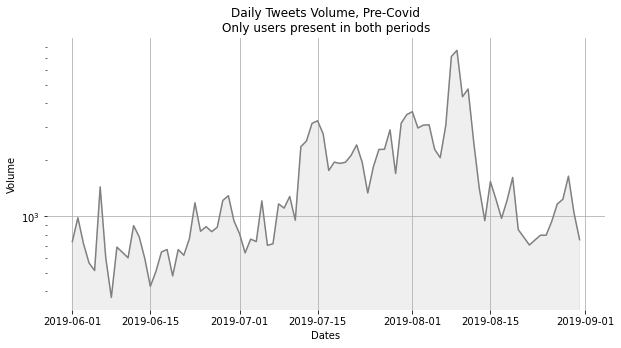}}
  {\label{f:tweets_post}
    \includegraphics[width=0.5\textwidth]{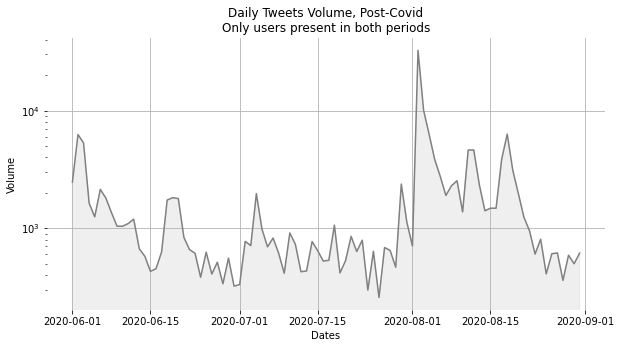}}
  
 \caption{Daily volume of tweets in the data set (on the left is the pre-pandemic period, and on the right is the post-pandemic period).}
 \label{f:tweets_ds}
\end{figure*}

Our dataset is composed of 892,487 tweets (192,087 are plain tweets, 700,400 are retweets --or RTs--, 109,097 are quotes, and 109,211 are replies) that talk about immigration in Chile, of which 299,098 correspond to a pre-pandemic period (June--August, 2019), and 593,389 to a post-pandemic period (June--August, 2020). Figure~\ref{f:tweets_ds} shows their daily volume. 
These tweets were written by 213,115 users (82,212 users analyzed in the pre-pandemic period and 130,903 in the post-pandemic period); 21,469 of these users are present in both analyzed periods, and our study focuses precisely on this subset of users since we can trace their attitude changes (if any). Hence, the number of tweets issued by the group of users present in both periods corresponds to 312,527 tweets: 146,743 tweets from the pre-pandemic period and 165,784 from the post-pandemic period.

\subsection{Classifier of Attitudes}

The methodology we applied has two steps, following the works of \textcite{freire2021framework, graells2022bots}: first, we identify the attitude of some users, i.e., we label them as empathetic (positive attitude) or threatened (negative attitude). Second, with the subset of users classified, we trained a classifier and predict the attitude of the rest of the dataset~\parencite{freund1999short, friedman2001greedy}. 

For the \textbf{labeling process}, we identified attitudes toward migration based on two social theories. The first one is the Intergroup Contact Theory, which proposes that contact between people from different groups will foster empathy and improve relationships~\parencite{allport1954nature}. The second perspective is the Integrated Threat Theory, which states that contact between people from different groups can cause prejudices and perceptions of threat, which will worsen relationships~\parencite{nelson2009handbook, stephan2013integrated}. Both theories tell us what to look for when analyzing attitudes toward migrants: either motivated by empathy, in favor of migration; or by threat, against migration. Thus, we will assume that there are two attitudes, which we labeled as \textit{empathy} and \textit{threat}, as in \textcite{freire2019characterization, freire2021framework}.

An effective mechanism to predict the community to which a user belongs is their choice of words~\parencite{bryden2013word}. Thus, we generated a list of seed patterns and keywords for each attitude. We iteratively explored the dataset in search of traits that could be assigned to these seeds and keywords. For empathetic attitudes, we focused on messages expressing that migration is a human right and that immigrants are welcome and deserve equal conditions (e.g., ``we are all immigrants'', ``no one is illegal'', etc.). On the other hand, for threatened attitudes, we focus on messages that express that immigrants are not welcome and/or negatively qualify their arrival or themselves (e.g., ``no more immigrants'', ``illegals take our jobs'', etc.).

\begin{table*}[t]
  \centering
  \begin{tabular}{|p{0.25\textwidth}|p{0.7\textwidth}|}
    \hline
    \textbf{Attitude}& \textbf{Patterns seeds and keywords} \\
    \hline
    \emph{Empathy} & \#bienvenidosachile, \#chilesinbarreras, \#chilediverso, \#noalaxenofobia, \#nomasracismo, \#bienvenidosmigrantes, \#nadieesilegal, \#todossomosmigrantes \ldots \\
    \hline
    \emph{Threat} & \#inmigrantesilegales, \#nomasinmigrantes, \#nomasilegales, \#vendepatria, \#fuerailegalesdechile, \#invasionmigrante, \#nomasinmigracionilegal, \#inmigracionilegaldesatada, invasi\'on \ldots \\
    \hline
  \end{tabular}
  \caption{Seed patterns and keywords for labeling each attitude toward immigration.}
  \label{tab:table_1}
\end{table*}

Once we obtained the final lists (see Table~\ref{tab:table_1}), we labeled the users that matched these patterns; in this way, we built our subset of training users for the classifier. In parallel to the labeling we did through the lists of seed patterns and keywords, we manually tagged some accounts of institutions (such as the International Office for Migration and the Jesuit Migrant Service), public figures, journalists, and politicians who have publicly expressed their attitude on migration.
This process was done for the data sets of both study periods, considering that during each period the users maintain only one attitude.

For the \textbf{classification process}, we use the subset of users labeled as training for the Extreme Gradient Boosting (XGBoost) classifier that trains decision trees using gradient boosting~\parencite{chen2016xgboost}.
The XGBoost classifier receives as an input a feature matrix defined as the concatenation of several matrices, three of them based on the content emitted by the users (such as a user-word matrix), and another three based on the interactions between them (such as an adjacency matrix of retweets). For more details see the Methodological Appendix~\ref{appendix:method}.  

Next, we trained the classifier using the set of labeled users to predict the attitudes of the remaining users (defined in the feature matrix).  
The classifier return a tuple $p$ for each user at each period, that we interpret as the  probabilities of belonging to each attitude. Each value from $p$ lies in $[0,1]$, and in our case, since we consider two attitudes, we can consider only of them as reference. We applied an uncertainty threshold of 0.05 to classify each user into one of those categories, i.e., $p$ must be either less than 0.45 (threat attitude) or greater than 0.55 (empathetic attitude). We defined users who could not be classified into either of the two attitudes as \emph{undisclosed}. 

Finally, we manually checked profiles that might have been mislabeled by the classifier, focusing on those that are very active/followed in the discussion. We added those manual labels and repeated the training/prediction steps until no obvious inconsistencies were found. 

\section{Results}
\label{sec:resultados}
The following section presents the main results of our analyses, beginning with a general description of the considered users and their attitudes towards migrants before and after the COVID-19 outbreak. Then, we analyzed both the reach and influence of empathetic and threatened users by observing their retweets in the same periods. Lastly, we performed a vocabulary analysis for both groups of users by attitude, exploring the pandemic-related words that each group mentioned more frequently in both periods. 

\subsection{General Description of Attitudes}

The total number of users analyzed is 231,115; 82,212 were analyzed in the pre-pandemic period, and 130,903 were analyzed in the post-pandemic period.
In the pre-pandemic period, 34,590 users were classified as empathetic, 16,141 as threatened, and 31,481 as undisclosed. On the other hand, in the post-pandemic period, 117,249 users were classified as empathetic, 11,678 as threatened, and 1,976 as undisclosed.

\begin{figure}[t]
  \subfloat{
   \label{fig:attit_dist_both}
    \includegraphics[width=0.5\linewidth]{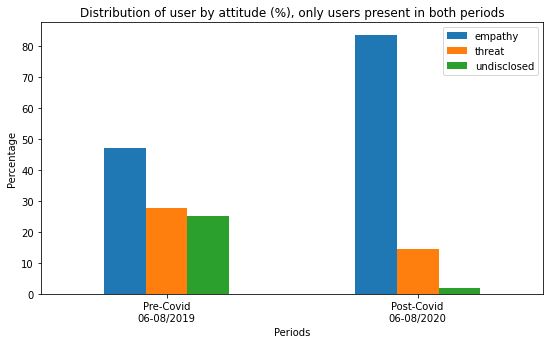}}
  {\label{fig:attit_dist_14}
    \includegraphics[width=0.5\linewidth]{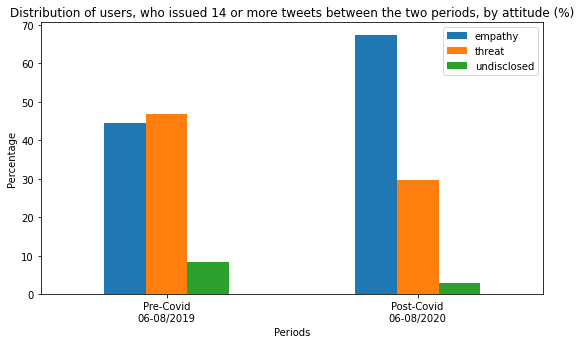}}  
 \caption{Percentage distribution of users, classified by attitude for each period of analysis. Left: users present in both periods ($n =$ 21.469). Right: users with more than 14 tweets in total, considering both period ($n =$ 5,312)}
 \label{fig:attit_dist}
\end{figure}

To study changes in attitude after the COVID-19 pandemic, we considered 21,469 users whose participation in the Twitter platform was observed in both periods, pre and post-pandemic. 
Figure~\ref{fig:attit_dist} (to the left) shows the distribution of users by attitude. In both periods, there were more empathetic users than threatened users, but pre-COVID-19 we found that less than 50\% of the users were classified as empathetic, while post-COVID-19 period nearly 80\% belonged to that category. Threatened users, as well as the undisclosed ones, decreased from one period to another. These results are different when looking only at users who tweet frequently, i.e., those that have more than 14 tweets about migration and migrants in total, considering both periods (5,312 users; 14 tweets is the average number of tweets considering both periods). In this subset of users, pre-COVID-19 threatened users were more numerous than empathetic ones; similarly, in the post-COVID-19 period the number of empathetic users increased, and the number of threatened and undisclosed users decreased, as we can see in Figure~\ref{fig:attit_dist} (to the right). 

\begin{figure}[t]
    \centering
    \includegraphics[width=0.6\linewidth]{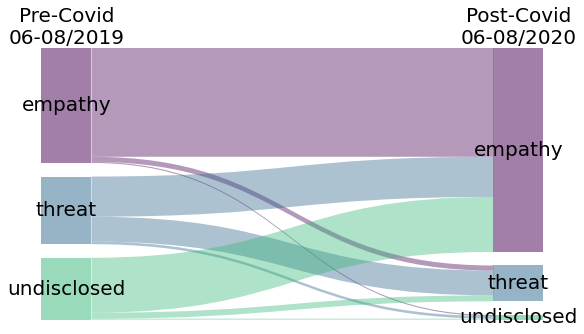}
    \caption{Number of users classified in each attitude, for the pre- and post-pandemic period, and their attitude changes.}
    \label{fig:change_atti}
\end{figure}

In addition, as we can see in Figure~\ref{fig:change_atti}, more than half of the users shifted their attitude from threatened to empathetic, and most of the undisclosed users in the pre-COVID-19 period shifted to empathetic in the post-COVID-19 period.

\begin{figure*}[t]
    \centering
    \includegraphics[width=\textwidth]{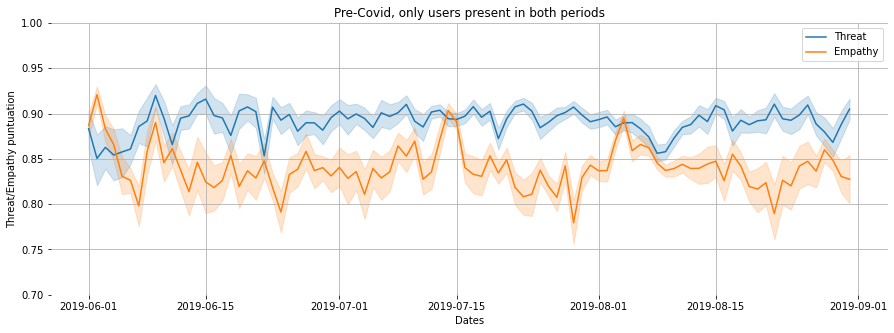}
    \includegraphics[width=1\textwidth]{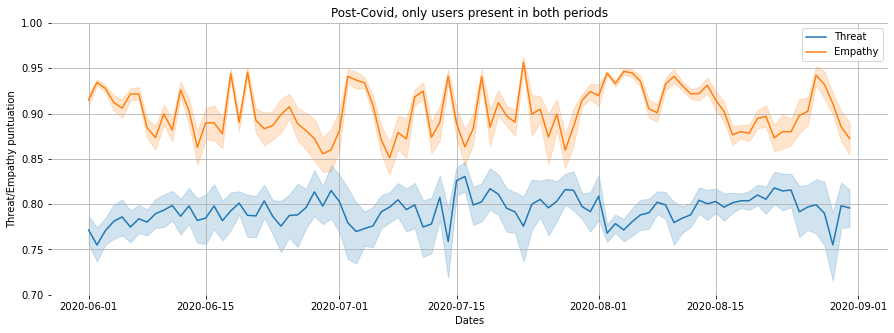}
    \caption{Daily distribution of average attitude scores (empathy and threat), for both periods. Top: pre-pandemic period. Bottom: post-pandemic period.}
    \label{fig:punt_dist}
\end{figure*}

Since threatened users decreased post-COVID-19, we explored the scores given to each user by the classifier, aiming to check if they increased between periods. In other words, we wanted to check if threatened and empathetic users were more or less threatened or empathetic before and after the COVID-19 outbreak. Figure~\ref{fig:punt_dist} shows the average daily score estimated using LOWESS: empathetic post-pandemic users scored more empathetic than the same group before the pandemic (from 0.78 to 0.91). Meanwhile, threatened users decreased their average scores from 0.87 to 0.80.

\subsection{Reach and Influence of Empathetic and Threatened Users}

Our first prediction stated that the number tweets by threatened users would increase (or, alternatively, that the number of tweets by empathetic users would decrease9. Tables~\ref{tab:table_2} and~\ref{tab:table_22} show the numbers of users, tweets, retweets (emitted and received), quotes, and replies by group and period. From this, we calculated the ratio of tweets by user, considering the users that were present in both periods. For empathetic users, the number of tweets per user was 5.4 pre-COVID-19 and 6.16 post-COVID-19. Meanwhile, for threatened users, the number of tweets per user was 13.6 pre-COVID-19 and 16.4 post-COVID-19. Thus, both kinds of users increased the number of tweets published post-pandemic, but for the threatened users the increase was greater. This result is consistent with the analysis of the complete dataset, where the increase is more prominent for the threatened group (from 9 to 15 compared to the empathetic group, which went from 3 to 3.4). In contrast, when we analyzed only users who have at least 14 tweets in total, although all users increase their tweets, the group of threatened users tweets the most and maintains similar average values in both periods (27.4 and 28.3). In addition, we observed a notorious increase in tweets per user for the empathetic group (from 14.4 to 19.5), and even more accentuated for the undisclosed group (from 5.5 to 20.7).

\begin{table*}[t]
 
  \centering
  \begin{tabularx}{\textwidth}{| X | X | X | X | X | X | X |}
    \hline
    Period & \multicolumn{6}{ |c| }{Pre-COVID-19} \\
    \hline
    Interaction/ Attitude & User & Tweet & Retweet emitted & Retweet received & Quote & Reply \\
    \hline
    Empathy & 10,100 & 54,597 & 42,047 & 46,511 & 6,460 & 4,602 \\
    \hline
    Threat & 5,936 & 80,883 & 61,131 & 83,722 & 10,447 & 12,191 \\ \hline
    Undisclosed & 5,433 & 11,263 & 7,144 & 7,630 & 1,039 & 2,374  \\ \hline
    Total & 21,469 & 146,743 & 110,322 & 137,863 & 17,946 & 19,167  \\ \hline
  
  \end{tabularx}
  
  \caption{Description of the number of users, tweets, retweets emitted and received, quotes, and replies issued by the different groups of users classified according to their attitude, for the pre-COVID-19 period.}
  \label{tab:table_2}
\end{table*}

\begin{table*}[t]
 
  \centering
  \begin{tabularx}{\textwidth}{| X | X | X | X | X | X | X |}
    \hline
    Period & \multicolumn{6}{ |c| }{Post-COVID-19} \\
    \hline
    Interaction/ Attitude & User & Tweet & Retweet emitted & Retweet received & Quote & Reply \\
    \hline
    Empathy & 17,906 & 110,235 & 92,967 & 156,688 & 12,760 & 7,515 \\ \hline
    Threat  & 3,147 & 51,439 & 42,520 & 51,654 & 7,460 & 5,103 \\ \hline
    Undisclosed & 416 & 4,110 & 2,877 & 7,526 & 505 & 750 \\ \hline
    Total & 21,469 & 165,784 & 138,364 & 215,868 & 20,725 & 13,368 \\ \hline
  \end{tabularx}
  \caption{Description of the number of users, tweets, retweets emitted and received, quotes, and replies issued by the different groups of users classified according to their attitude, for the post-COVID-19 period.}
  \label{tab:table_22}
\end{table*}

This only informed us of users’ activity and interactions with others in this environment, not the influence or reach that they have. To check the influence of each kind of user before and after the COVID-19 outbreak, we calculated the ratio of retweets per tweet by attitude for both periods (dividing the number of retweets received by each user group by the number of tweets, issued by each user group). The results for threatened users showed a higher ratio before the pandemic (1.04 retweets per tweet), compared to empathetic users (0.85). Also, the empathetics’ ratio increased after the pandemics (1.42), a value still greater than for threatened ones (1.0); the latter remaining stable between both periods. Analysis of the full dataset shows that all users slightly increased their average number of retweets; in contrast to this, for the group of users with at least 14 tweets, only the threatened group showed a decrease in their retweets per tweet between the pre- and post-pandemic period, while the empathetic and undisclosed increased their ratios. This measure considered that there may be users that got more retweets than others because they also tweeted more frequently, so it gave us information on the scope and spread of information issued by each user group according to their attitudes.

\subsection{Vocabulary Analysis}

As a final qualitative analysis, we explored differences in language use. To do so, we calculated the proportional change between the words most used by empathetic and threatened users. We considered the 50 most frequent words in pre-pandemic and the 50 most frequent words in post-pandemic period. 

\begin{figure*}[htp]
    \centering
    \includegraphics[width=\linewidth]{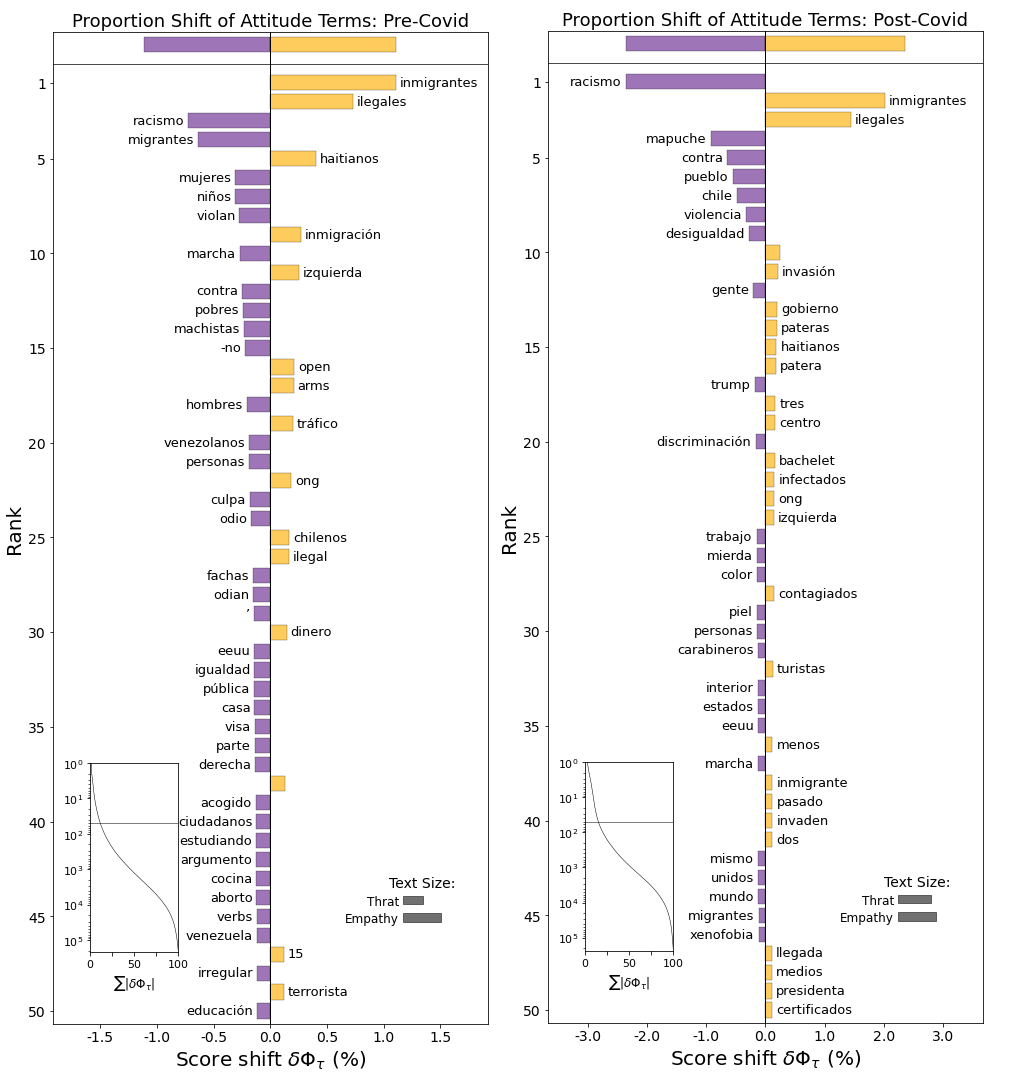}
    \caption{Top-50 words with relative frequency per attitude, pre- and post-pandemic (left and right, respectively).}
    \label{fig:shift}
\end{figure*}

Figure~\ref{fig:shift} shows the differences in the relative frequency of word use between the two groups of users. Words used in greater proportion by the empathethic group are shown in purple, and orange otherwise. Here we observe that in both periods the empathetic group used words such as ``children (ni\~nos)'', ``women (mujeres)'', ``men (hombres)'', ``persons (personas)'' and ``people (gente)''; which categorize immigrants in a neutral and equal manner; in contrast, the words ``illegal (ilegales)'', ``tourists (turistas)'' and ``invasion/invade (invasi\'on/invaden)'' were used by threatened users. Furthermore, the empathethic group used words such as ``equality (igualdad)'', ``welcomed (acogido)'', ``education (educaci\'on)'' in the pre-pandemic period, denoting their perception of a good reception towards immigrants; in contrast, for the post-pandemic period, the words become more accusatory against threatened users who do not place migrants in an equal status as locals (with words such as ``inequality (desigualdad)'', ``discrimination (discriminaci\'on)'' and ``xenophobia (xenofobia)'').

Finally, we note that the group of threatened users expressed their threat perceptions around delinquency in the pre-pandemic period, with words such as ``arms (armas)'', ``trafficking (tr\'afico)'', ``terrorists (terroristas)''. Such threat perceptions in the post-pandemic period were replaced by threats of contagion by COVID-19, as their prominent words were related to the pandemic, such as ``infected (infectados)'' and ``contagious (contagiados)''.

\begin{figure*}[ht]
    \centering
    \includegraphics[width=\textwidth]{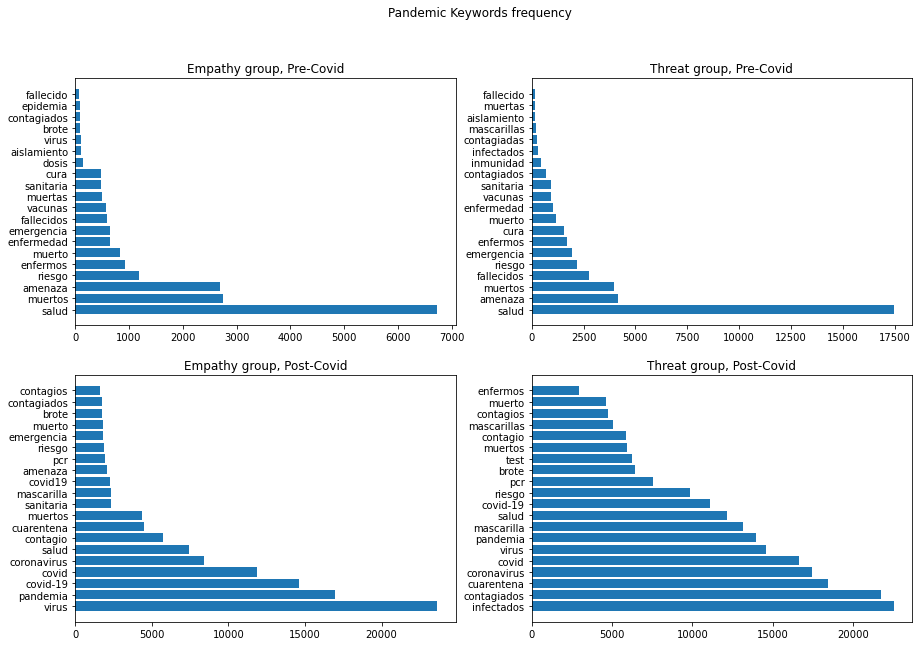}
    \caption{Use of pandemic-related keywords in each attitude group, pre and post-COVID-19.}
    \label{fig:vocab}
\end{figure*}

To study the change in pathogen threat perception in both attitude groups, we explored the pandemic-related keywords (152) that each group of users mentioned, for both periods. Figure~\ref{fig:vocab} shows that in the post COVID-19 period, the empathetic group used words related to the disease itself, such as ``virus'', ``pandemic (pandemia)'', and ``COVID-19'', while threatened users used words with a more personal connotation, such as ``contagiados'' o ``infectados'' (both meaning ``infected'').

\section{Discussion}
\label{sec:discu}
One of the core statements of the Behavioral Immune System (BIS) is that people tend to enhance their rejection of minorities and foreign groups under the threat of contagious diseases~\parencite{schaller2011behavioral}. This rejection would be motivated by the activation of the BIS and the consequent constriction in sociality. Thus, this study aimed to test whether threatened attitudes towards immigrants changed with the COVID-19 outbreak, predicting that these attitudes would increase after the pandemic. Our results partially support this hypothesis.

Specifically, we expected an increase in the number of Twitter users with threatened attitudes or a decrease in the number of users with empathetic attitudes, between pre- and post-pandemic contexts. Our results show that, in both periods, empathetic are more than threatened users. Threatened users, as well as the undisclosed ones, decreased from one period to another. The undisclosed group stands out for having a very large decrease: this indicates that in the post-pandemic period, it was easier to identify user attitudes for our classifier, which can be interpreted as users becoming clearer in their positions and making more noticeable their trends in attitudes. However, these results differ when we focused on vocal users with more than 14 tweets about migration and migrants in total (average number of tweets users have between the pre- and post-pandemic period), considering both periods (5,312 users): here, pre-COVID-19 threatened users were more numerous than empathetic ones. This result is interesting as generally, the population with positive attitudes is higher than those with negative attitudes~\parencite{pettigrew2006meta, freire2019characterization, freire2021framework}. Given that this group of users shows greater participation in the migration debate and holds more extreme attitudes and positions (which is reflected in the low proportion of users classified as undisclosed), we can see that the proportion of highly participative users with negative attitudes towards the migration debate is greater than those with positive attitudes. Given that in the post-COVID-19 period this phenomenon is reversed (the number of empathetic users increases, becoming greater than those threatened) and a similar result is recovered as when we look at all the users analyzed, we could infer that the pandemic dampened the negative attitudes and increased the positive ones for the group of highly participative users.

Similarly, we expected that the number of tweets by threatened users would increase or the number of tweets by empathetic users would decrease. The evidence for this is that, for both attitudes, the number of tweets per user increased between periods. However, threatened users have a higher number of tweets per user in both periods. Furthermore, although both types of users increased the number of tweets they posted after the pandemic, for threatened users the increase was greater. Thus, we can interpret that the group of users with negative attitudes produces more content than the group with positive attitudes, and they extreme their positions in the post-pandemic period, which supports the BIS hypothesis. The result is different when we look at the group of highly active users (who have 14 or more tweets in both periods), since the average production by threatened users remains similar (from 27.4 to 28.3 tweets per user), and the empathetic and undisclosed groups show the greatest increase, the latter standing out with an increase from 5.5 to 20.1 tweets per user. This implies that the threatened highly active group shows more stable behavior, and it is the empathetics that show the greatest change in the face of the pandemic.

Although these results seem to partially go against the BIS hypothesis, we must consider that a large part of the migrant population worked in essential services (health professionals, supermarkets, pharmacies, garbage collection, deliveries, etc.) during the COVID-19 pandemic; therefore, the local population mostly maintained contact with them (in urban contexts)~\parencite{fasani2020immigrant, gelatt2020immigrant}. Thus, we could interpret that this contact favored the positive perception of migrants~\parencite{poole2023expedient}, and with it, the increase of empathetic users and the decrease of threatened users. With this in mind, the Inter-group Contact Theory may explain this result~\parencite{allport1954nature}. The main predictions of this theoretical model are that increased contact between locals and immigrants will promote empathy and understanding between them, and consequently improve and increase positive attitudes towards immigrants.

We also expected an increase in the number of retweets from the group of users with threatened attitudes between pre- and post-pandemic contexts or a decrease in the number of retweets from the empathetic group of users. To check the influence of each attitude before and after the COVID-19 outbreak, we calculated the ratio of retweets per tweet by attitude for both periods. Threatened users had a greater ratio before the pandemic, compared to empathetic users, meaning that their reach was higher in that period. However, the empathics' ratio increased after the pandemic, remaining greater than the threatened's ratio, the latter remaining stable between both periods. This measure considered that there may be users that got more retweets than others because they also tweeted more frequently, so it gives us information on the scope and spread of information issued by each user group according to their attitudes.

Lastly, we expected that threatened users would use COVID-related words more frequently than empathetic users in the post-pandemic context. We verified this prediction in the first instance since the threatened group is the only one that uses words related to the pandemic (such as ``infected'') in the post-pandemic period. In addition, we found that in the pre-pandemic period, this group of users showed threat perceptions of crime (using words such as ``arm'', ``terrorist'' and ``trafficking''), so the COVID-19 outbreak may have had an effect on the change of threat perception from crime to pathogen contagion.

Zooming in on the language related to the pandemic, we explored the use of 152 pandemic-related keywords for each group of users. Our results point out that the empathetic group used words related to the disease itself, like ``virus'', ``pandemia'', and ``COVID-19'', while the threatened group tended to use words focused on individuals who posed a threat of contagion, similar to what was found by \textcite{freire2019characterization}, where the empathetic group of users referred to migration as a general phenomenon, and the threatened group referred to the particular phenomenon that concerned them. In addition, the total number of pandemic keywords used by the threatened group is higher than the empathetic group for both periods. Interestingly, when we adjusted for the number of users for each group and period, we saw that the use of these words per user decreased for empathetic users from 256 to 126 between periods, but it increased for the threatened group (from 1035 to 1462 post COVID-19 outbreak). This would reflect the fact that for this group of users, the perception of a threat of infection from the interaction with the immigrant population is higher. 

Thus, we have not found clear evidence of increased negative attitudes towards migrants, similar to the findings of \textcite{rowe2021using, schwambergova2023pandemic}. The fact that we have no validation of the hypotheses derived from the BIS is certainly influenced by a large number of factors internal and external to the population under study.
We can mention, for example, that anti-immigrant sentiment was not established in Chile because it is a country that, although it has historically received migrants, its migrant population was very small until seven years ago, at which time Chile experienced a drastic increase in immigration~\parencite{Cepal}. In addition, the migrant population in Chile is mostly Latin American \parencite{Cepal}, and this particular population was not directly associated with COVID-19, as was the Asian population, who were the focus of discrimination in different parts of the world~\parencite{organisation2020impact, coates2020covid}. Accordingly, it is possible that the immigrant population was seen by most of the Chilean population as helpers or victims rather than a threat. 

Our study acknowledges certain constraints that may affect the generalization of our findings. Twitter data, while rich, may not be fully representative of the entire population, as it is biased towards certain demographics that have access to and engage with this platform. Such biases may reflect regional variations and technological accessibility, yielding a potentially skewed picture of public opinion on immigration. Additionally, our methodology assumed temporal stability in attitudes by averaging user sentiments across tweets, potentially overlooking more nuanced, real-time shifts in public opinion. This is particularly pertinent given that attitudes towards immigrants can be significantly swayed by temporal events such as news cycles or political developments~\parencite{freire2019characterization}. Moreover, our analysis is limited by the absence of additional data to contextualize user attitudes, such as political affiliation, socio-economic status, or opinions on related subjects. The thematic filtering necessary for our analysis further narrows the scope, excluding potentially pertinent topics that co-occur with migration discourse. Finally, the inherent privacy policies of the platform limit the availability of more detailed user demographics that could enrich the understanding of the factors shaping public attitudes.

Two interesting lines of research emerge from our work; firstly, a temporal analysis of attitudes towards migration in conjunction with a detailed analysis of the most relevant pandemic events in the period studied, and their possible influence on the formation and/or change of these attitudes, and secondly, a geo-referenced analysis of users, to cross-reference socio-demographic information such as the national census \parencite{graells2020representativeness} and our measurement of attitudes towards migration.

Despite these limitations, this work presents the novelty of using Twitter to classify and then analyze attitudes towards migration, comparing the pre- and post-pandemic context. This platform allows to collect a large volume of data at a relatively low cost and it is also capable of capturing the spontaneous manifestation of different points of view on a given topic.

\section{Conclusions}
\label{sec:conclu}
The objective of this study was to test whether threat attitudes toward immigrants changed with the COVID-19 outbreak; predicting, according to one of the BIS hypotheses, that these attitudes would increase after the pandemic. The evidence does not go in one direction. On the one hand, in support of the BIS hypothesis, threatened users increased their tweet production in the post-pandemic period in greater quantity than empathetic users. While empathetic users increased the reach of their tweets between the two periods, the ratio of retweets per user increased, remaining over the ratio calculated for threatened users. In addition, only threatened users showed a change in threat perception between the pre- and post-pandemic periods, moving from a concern about crime to one about COVID-19 infection. In contrast, empathetic users employed words related to the disease itself, whereas threatened users tended to employ words with a more personal connotation; making it noticeable that users with negative attitudes towards migration showed perceptions of threat.

It is important to note that threatened users were found to decrease from period to period, with the undisclosed group standing out as having a very large decrease, indicating that users became clearer in their positions and made their attitudinal tendencies more noticeable. However, when considering highly active users, pre-COVID-19 threatened users were more numerous than empathetic users. The pandemic attenuated the negative attitudes and increased the positive ones for the highly active user group. The Intergroup Contact Theory may explain this result, given that most of the services continued to be provided by immigrants, so that they had to face greater exposure to the virus, thus generating empathy in the local population.

Finally, we can infer that the COVID-19 pandemic may have increased the group of users with positive attitudes on a general level, just as it made the group with negative attitudes more vociferous.
This work is a contribution to the studies of the impact of the COVID-19 pandemic in Chile. In addition, it addresses the migration phenomenon and the perceptions that arise from it, and it shows that it is possible to test theoretical models that predict the attitudes, feelings and concerns of the Chilean population towards immigration in the pandemic context.

\printbibliography

\if@endfloat\clearpage\processdelayedfloats\clearpage\fi 

\appendix
\section{Appendix}
\subsection{Keywords used for filtering}
\label{appendix:keywords}

\begin{itemize}
    \item inmigraci\'on 
    \item inmigrante
    \item migraci\'on
    \item migrante
    \item interculturalidad
    \item xenofobia
    \item xenof\'obo/xen\'ofoba
    \item xenof\'obico/xenof\'obica
    \item racismo
    \item racista
    \item promigrante
    \item \#migraci\'on
    \item \#migranteschile
    \item \#todossomosmigrantes
    \item \#redmigrante
    \item \#migracionpdi
    \item \#chilesinbarreras
    \item \#migraci\'onlaboral
    \item \#leydemigraci\'on
    \item \#leydemigraciones
    \item \#leymigratoriaya
    \item \#reformamigratoria
    \item \#chileterecibe
    \item \#aricaabrefronteras
    \item \#bajoelmismosol
    \item \#interculturalidad
    \item \#encuentromigrante
    \item \#mipieltupiel
    \item \#servicioevangelicomigrantes
    \item florvil
    \item @oimchile
    \item \#inmigrantesilegales
    \item \#nomasinmigrantes
    \item \#nosmasilegales
    \item \#justiciaparajoane  
    \item \#construyendocultura
    \item \#multicolor
    \item \#nomasdiscriminacion
    \item \#sinfronteras
    \item \#multiculturalismo
    \item \#conlosrefugiados
    \item \#nohayserhumanoilegal
    \item \#multicultural
    \item \#migrafest 
    \item \#bienvenidosnuevoschilenos
    \item charter
    \item @dptoextranjeria
    \item @oimchile
    \item @somosfre
    \item @sjmchile
    \item \#chileterecibe
    \item \#lepraenchile
    \item \#migracionysalud
\end{itemize}

\subsection{Methodological details of XGBoost}
\label{appendix:method}

The XGBoost algorithm is based on the construction of a model that is a weighted combination of several simpler models. Each simpler model is a decision tree and is called a ``weak tree''.

Given a training set $D={(x_{1},y_{1}),\dots,(x_{n},y_{n})}$ with $n$ instances and $y_{i}\in[0,1]$ for binary classification, XGBoost searches for a prediction function $F(x)$ that minimizes a loss function $L(y,F(x))$.

The XGBoost algorithm uses the boosting technique, which consists of building sequential models in which each model attempts to correct the errors of the previous model \parencite{freund1999short}. To do so, at each iteration $t$, a new weak tree $f_t(x)$ is trained to fit the residuals $r_{i,t}$ of the current model $F_{t-1}(x_i)$, i.e., $r_{i,t}=y_{i}-F_{t-1}(x_{i})$.

The goal is to minimize the loss function at each iteration:
\begin{equation}
    L^{(t)} = \sum_{i=1}^{n} L(y_{i}, F_{t-1}(x_{i}+f_{t}(x_{i}))+\Omega(f_{t}),
\end{equation}
where $L(y_{i},F_{t-1}(x_i)+f_{t}(x_i))$ is the loss function corresponding to the addition of the new tree $f_{t}$ to the current model $F_{t-1}$, and $\Omega(f_{t})$ is a regularization function that penalizes the complexity of the tree $f_{t}$ to avoid overfitting.

To find the optimal weak tree $f_{t}$, XGBoost uses a technique known as "split finding", which looks for the best way to split the data into two groups to minimize the loss function. At each node of the tree, all possible splits of the data are considered according to an input variable and a cut-off point, and the split that minimizes the loss function is chosen.

Once the optimal weak tree $f_{t}$ has been found, a learning factor $\eta$ is adjusted to control the contribution of this tree to the final model. Finally, the weighted weak tree is added to the current model to obtain the new model:

\begin{equation}
    F_{t}(x)=F_{t-1}(x)+\eta f_{t}(x).
\end{equation}

This process is repeated until a maximum number of iterations is reached or until an acceptable model accuracy is reached.

For our study, the XGBoost  classifier receives as an input a feature matrix defined as the concatenation of several matrices, three of them based on the content emitted by the users, and another three based on the interactions between them:

\begin{itemize}
    \item In the content-based matrix each row represents user $i$, and each term $j$ can represent a word, hashtag, username, URL or emoji. Thus, a cell $(i, j)$ contains the number of times user $i$ has used the term $j$ in his tweets, in his biographical self-description and in his profile URL.
    \item For the interactions matrix we consider that homophily can vary or be absent in different layers of interaction ~\parencite{manivannan2018different}, so we use the three types of interaction allowed by the platform to build it; retweets, replies and quotes. Thus, the matrix stores in a cell $(i, j)$, the number of times user $i$ has interacted with user $j$ (e.g., if $i$ retweets $j$ once, $c(i,j) = 1$). In addition, for each type of interaction, we consider whether the user has interacted with other users already labeled with an attitude.
\end{itemize}

Thus, we train the classifier using the set of labeled users to predict the attitudes of the remaining users (defined in the matrix described above). As parameters of the classifier that allow us to avoid overfitting; first, gradient boosting is performed with an early stop, using a validation set of 10\% of the training observations. Second, we remove from the feature matrix the seed keywords we used in the training labeling, since our goal is to classify users that do not use these terms in their content. 

The classifier gives us for each user $u$ a value $p_a(u)$, which corresponds to the fraction of decision trees that vote for the corresponding attitude $a$. This value is at $[0,1]$, and in our case, since we consider two attitudes, we define the score of one attitude as a function of the other $p_{\text{empathy}}(u) = 1 - p_{\text{threat}}(u)$. Then, we apply a threshold ($p>0.55$) to consider predictions with a number of voters higher than a random choice. We define users who could not be classified into either of the two attitudes as \emph{undisclosed}. For our analyses, we used both $p$ and categorical classification of each user and tweet.

Finally, we manually check profiles that might have been mislabeled by the classifier, focusing on those that are very active/followed in the discussion. We add those manual labels and repeat this step until no obvious inconsistencies are found.

\end{document}